\documentclass[showpacs,preprintnumbers,amsmath,amssymb,pre]{revtex4}
\usepackage{amsmath}
\usepackage{graphicx}
\usepackage{bm}% bold math

\begin{document}

\title{On the rise and fall of networked societies}

\author{George C.M.A. Ehrhardt}
\email{gehrhard@ictp.trieste.it}
\author{Matteo Marsili}
\affiliation{The Abdus Salam International Centre for Theoretical Physics, Strada Costiera 11, 34014 Trieste Italy}
\author{Fernando Vega-Redondo}
\affiliation{Departamento de Fundamentos del An\'{a}lisis Econ\'{o}mico and Instituto Valenciano de Investigaciones Econ\'{o}micas, Universidad de Alicante, 03071 Alicante, Spain}

\begin{abstract}
We review recent results on the dynamics of social networks which  suggest
that the interplay between the network formation process and  volatility may
lead to the occurrence of discontinuous phase  transitions and phase
coexistence in a large class of models.  We then investigate the effects of
negative links -- links inhibiting  local growth of the network -- and of a
geographical distribution of  the agents in such models. We show, by
extensive numerical  simulations, that both effects enhance this
phenomenology, i.e. it  increases the size of the coexistence region.
\end{abstract}


\maketitle

%PACS numbers:  89.75.Hc, 02.50.Ey.

%%%%%%%%%%%%%%%%%%%%%%%%%%%%%%%%%%%%%%%%%%%%
%% MAINMATTER
%%%%%%%%%%%%%%%%%%%%%%%%%%%%%%%%%%%%%%%%%%%%

\section{Introduction}

Recent phenomenological studies on complex networks in the social sciences have
uncovered ubiquitous nontrivial statistical properties, such as scale free
distribution of connectivity or small world phenomena \cite%
{alb_bar_01,dor_men_01,wa_stro_98}. These properties have striking
consequences on the processes which take place on such networks, such as
percolation \cite{dor_men_sam_01a}, diffusion \cite%
{far_der_bar_vic_01,goh_kah_kim_01b}, phase transitions \cite{DGM,EM} and
epidemic spreading \cite{vespignani}. The research on complex networks
raises questions of a new type as it addresses phenomena where the topology
of interactions is part of the dynamic process. This contrasts with
traditional statistical mechanics, where the topology of the interaction is
fixed \emph{a priori} by the topology of the embedding space.

Phenomena of this type are quite common in social sciences where agents
purposefully establish cooperative links \cite{Putnam}. Links between
individuals in a social network support not only the socioeconomic
interactions that determine their payoffs, but also carry information
about the state of the network. This aspect has important consequences in
the long run if the underlying environment is volatile. In this case, former
choices tend to become obsolete and individuals must swiftly search for new
opportunities to offset negative events. The role of the network for
information diffusion is particularly apparent, for example, pertaining to
the way in which individuals find new job opportunities. For example, it has
been consistently shown by sociologists and economists alike \cite{G,T} that
personal acquaintances play a prominent role in job search. This, in turn,
leads to a significant correlation in employment across friends, relatives, or
neighbours. The common thesis proposed to explain this evidence is that, in
the presence of environmental volatility, the quantity and quality of one's
social links -- sometimes referred to as her \emph{social capital} \cite%
{Coleman}-- is a key basis for search and adaptability to change.

A recent statistical mechanics approach to simple models of social networks
has recently shown that the interplay between volatility and the quest for
efficiency leads, in a broad class of models, to a positive feedback loop
between the network's structure and its dynamics \cite{MSV,EMV}. As a
result, social networks may exhibit sharp phase transitions -- i.e. a dense
network may emerge or disappear \cite{kolkata} suddenly -- coexistence of
different network phases for the same parameters and resilience -- i.e.
robustness of a dense social network even when external conditions
deteriorate beyond the point where a dense network first came into
existence. This generic conclusion was derived in two qualitatively
different setups: Ref. \cite{MSV} addressed the interplay between volatility
and search in a model where agents use their links to look for new fruitful
collaborations. Ref. \cite{EMV} found instead the same phenomenology in
generic models where proximity or similarity favours the formation of links
among agents and, conversely, the presence of links between two agents
enhances similarity. As discussed in Ref. \cite{EMV}, there are several
socio-economic phenomena, ranging from job contact networks and research
collaborations to the spread of crime and other social pathologies, for
which anecdotal evidence has been reported.

Such dynamic effects (e.g. sharp transitions) are much more difficult to
detect in empirical studies than static properties (e.g. scale-freeness or
small-worldness). Hence, the empirical verification of the scenarios
proposed in Refs. \cite{MSV,EMV} requires very accurate data, which is
rarely available in socio-economic sciences. A different way to check the
validity of the scenario in generic cases is to challenge its robustness on
theoretical grounds, including effects which have been neglected so far.
Here, in particular, we address two simplifying assumptions of the models of
Ref. \cite{MSV,EMV}: \emph{i)} that network formation is long ranged, i.e.
independent of a geographical distribution of the agents and \emph{ii)} that
links always have a positive effect on the link formation process.

We discuss these effects in the framework of the model of Ref. \cite{MSV}
where they enter in an important way into the dynamics of the network (see
later). In both cases, we find by extensive numerical simulations, that
inclusion of these effects enhances the character of our conclusions (i.e.
it increases the co-existence region in parameter space). This supports the
conclusion that sharp transitions, co-existence and resilience are generic
dynamic properties of social networks.

In what follows we shall first review the model of Ref. \cite{MSV}, then
turn to the study of negative links and finally discuss the inclusion of
geographical effects.

\section{Searching partners in a volatile world}

Ref. \cite{MSV} proposes a stylized model of a society that embodies the
following three features: (i) agent interaction, (ii) search and (iii)
volatility (i.e. random link removal). Individuals are involved in bilateral
interactions, as reflected by the prevailing network. Through occasional
update, some of the existing links have their value deteriorate and are
therefore lost. In contrast, the individuals also receive opportunities to
search that, when successful, allow the establishment of fresh new links.

Formally, the network is given by a set of nodes $N$ and the corresponding
adjacency matrix $A(t)$ with elements $a_{ij}(t)=1$ if there is a link
connecting nodes $i$ and $j$ at time $t$, and $a_{ij}=0$ otherwise (we
assume no on-site loops, $a_{ii}=0$ and un-oriented links $a_{ij}=a_{ji}$).
Denote by $F_{i}=\{j|a_{ij}=1\}$ the set of neighbours (\textquotedblleft
friends\textquotedblright ) of the node $i$. The matrix $A(t)$ follows a
stochastic process governed by the following three processes:

\begin{description}
\item \emph{Long distance search}: At rate $\eta $, each node $i$ gets the
opportunity to make a link to a node $j$ randomly selected (if the link is
already there nothing happens).

\item \emph{Short distance search}: At rate $\xi ,$ each node $i$ picks at
random one of its neighbours $j\in F_{i}$ and $j$ then randomly selects
(i.e. \textquotedblleft refers to\textquotedblright ) one of its other
neighbours $k\in F_{j}\backslash \{i\}$. If $k\not\in F_{i}$ then the link
between $i$ and $k$ is formed. If $F_{i}=\emptyset $ or $F_{j}=\{i\}$ or $%
k\in F_{i}$ nothing happens.

\item \emph{Decay}: At rate $\lambda ,$ each existing link decays and it is
randomly deleted.
\end{description}

\begin{figure}[tbph]
\includegraphics[width=90mm,clip]{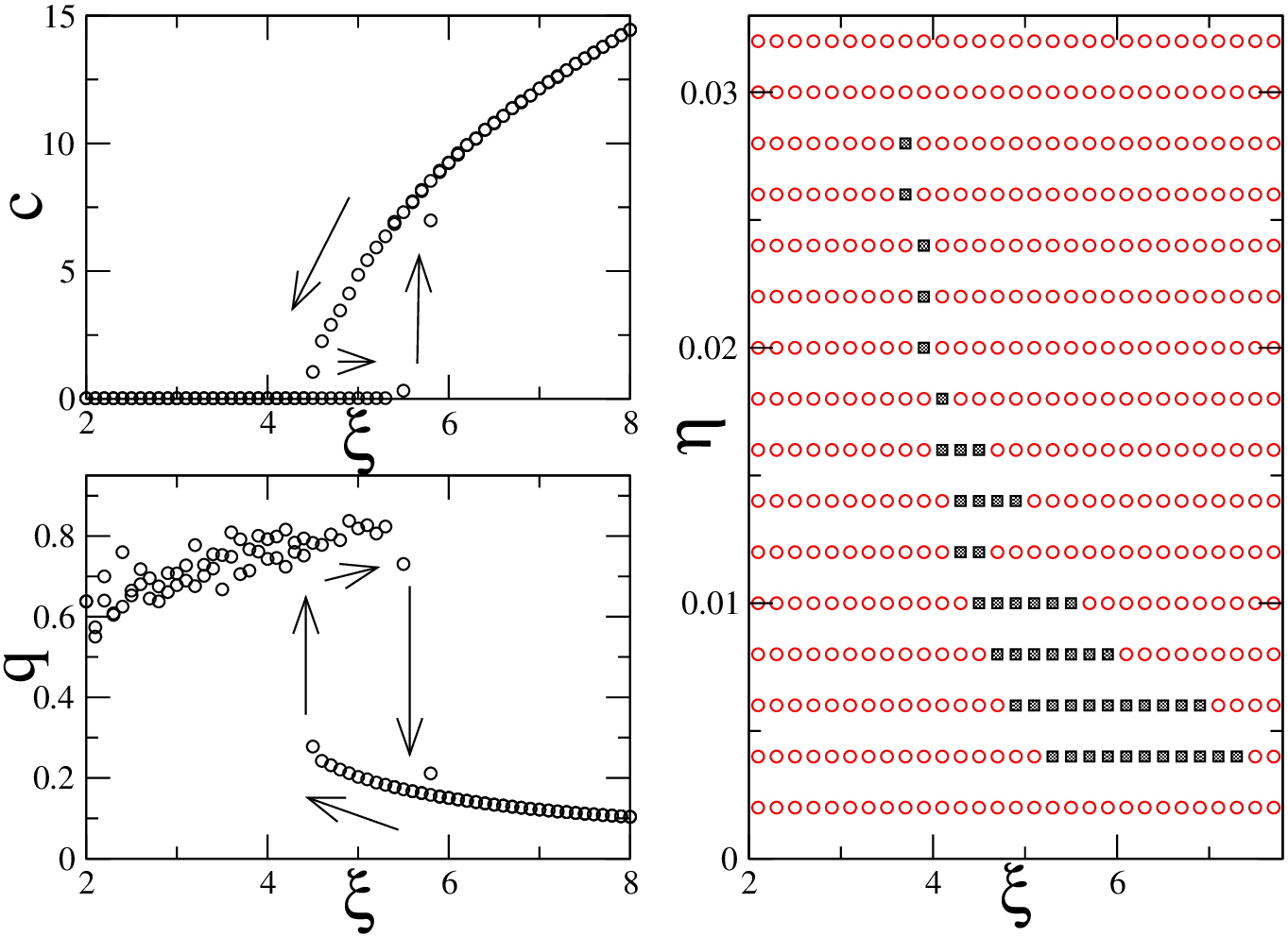}
\caption{Average degree $c$ (top) and clustering coefficient $q$ (bottom)
from numerical simulations with $\protect\eta /\protect\lambda =0.01$ for
populations of size $n=1000$. Here and in all other figures, runs were
equilibrated for a time $t_{\mathrm{eq}}=3000/\protect\lambda $ before
taking averages for a further $3000/\protect\lambda$. The network was
started in both the low connected and high connected state for each value of 
$\protect\xi$. For the central coexistence region, the two distinct points
for each $\protect\xi$ represent the two different starting configurations.
The arrows show the hysteretic region, the rightmost arrows indicating $\xi_2$.
The right hand graph shows the phase diagram, black squares denote the
coexistence region, red circles the regions in which only the low (lower
left) or high (upper right) phases are stable. }
\label{fig1_3}%\vspace*{60mm} %
\end{figure}

Over time, this process leads to an evolving social network that is always
adapting to changing conditions. For $\xi =0,$ the dynamics is very simple
and the stationary network is a random graph with average degree $c=2\eta
/\lambda $. For $\eta \ll \lambda $ the network is composed of many
disconnected parts. Fig. \ref{fig1_3} reports what happens when the local
search rate $\xi $ is turned on. For small $\xi ,$ network growth is limited
by the global search process that proceeds at rate $\eta $. Clusters of more
than $2$ nodes are rare and, when they form, local search quickly saturates
the possibilities of forming new links. Suddenly, at a critical value $\xi
_{2}$, a giant component connecting a finite fraction of the nodes emerges.
The average degree $c$ indeed jumps abruptly at $\xi _{2}$. The distribution 
$p(c)$ of $c_{i}$ is peaked with an exponential decrease for large $c$ and a
power law $p(c)\sim c^{\mu }$ for $c$ small. The network becomes more and
more densely connected as $\xi $ increases further. But when $\xi $
decreases, we observe that the giant component remains stable also beyond
the transition point ($\xi <\xi _{2}$). Only at a second point $\xi _{1}$
does the network lose stability and the population gets back to an
unconnected state. There is a whole interval $[\xi _{1},\xi _{2}]$ where
both a dense-network phase and one with a nearly empty network coexist. This
behaviour is typical of first-order phase transitions. The coexistence
region $[\xi _{1},\xi _{2}]$ shrinks as $\eta $ increases. 
%and it disappears for $\eta >0.05\lambda $. HM, only for theory.

In loose words, the model shows that the continuous struggle of agents'
continuous search must be strong enough to offset volatility if a dense and
effective social network is to be preserved. On the other hand, search can
be effective only in a densely networked society. So information diffusion
and a dense network of interactions are two elements of a feedback
self-reinforcing loop. As a result, the system displays a discontinuous
phase transition and hysteresis, enjoying some resistance to a moderate
deterioration of the underlying environmental conditions. Such a resilience
can be interpreted as consequence of the buffer effects and enhanced
flexibility enjoyed by a society that has accumulated high levels of social
capital.

These features are captured by a mean field theory which is in good
qualitative agreement with numerical simulation results (see Ref. \cite{MSV}%
). This theory highlights the particular role that clustering plays in the
dynamics of the model. Indeed search is particularly effective when
clustering is low whereas it is suppressed in a high clustered society. The
average clustering coefficient $q$ -- defined as the fraction of pairs of
neighbours of $i$ who are also neighbours among themselves\footnote{%
The averaging is done only over nodes with at least two neighbours.} --
shows a non-trivial behaviour. In the unconnected phase, $q$ increases with $%
\xi $ as expected. In this phase, $q$ is close to one because the expansion
of the network is mostly carried out through global search, and local search
quickly saturates all possibilities of new connections. On the other hand,
in the dense-network phase, $q$ takes relatively small values. This makes
local search very effective. Remarkably we find that $q$ \emph{decreases}
with $\xi $ in this phase, which is rather counterintuitive: \emph{%
increasing the rate }$\xi $\emph{\ at which bonds between neighbours form
through local search, the density $q$ of these bonds decreases}. In fact,
similar behaviour is found if, fixing $\xi $ and $\eta ,$ the volatility
rate $\lambda $ decreases.

These conclusions rest on two basic assumptions, which might be unrealistic
in practical cases. The first is that links have always a positive effect on
the formation of other links. Indeed, \textquotedblleft
negative\textquotedblright\ links (i.e. animosity) may have an important
effect in inhibiting link formation. If one of the my possible friends has a
negative relation with a friend of mine, I might not wish to form the link
with him/her, because this would increase the \textquotedblleft
frustration\textquotedblright\ of my social neighbourhood. It is indeed a
well accepted fact in social science \cite{balance} that social
relationships evolve in such a way as to decrease frustration.

The second assumption of the model, is that agents are treated equivalently
in the global search process. In many real cases, agents are located in a
geometrical space and this influences the likelihood with which they
establish new links among themselves. Notice that a dependence of the link
formation rate on proximity in space has arguably strong consequences on
clustering, which is a key aspect of the model.

In both cases, as we shall see, the inclusion of these effects enhances the
non-linear effect and result in an even wider region of coexistence.

\section{The effect of negative links}

Here we extend the model to include the effect of negative links. In
addition to the long-range search, introduction of friends, and decay of
links, we also include negative links. These links model the effect of
animosity between nodes. Thus, when two nodes $i$ and $j$ are introduced,
before they form a positive link they check through all their neighbours to
see if any of them have a negative link with their prospective neighbour.
They are in effect using their contacts to check the 'references' of their
prospective neighbour. If there are one or more negative links (or if $i$
and $j$ already have a negative link) then the new connection is not formed.

Negative links themselves are formed by the `souring' of existing positive
links at a rate $\gamma $. In other words, every link is positive when it is
created, but it may turn to negative at rate $\gamma $. Once formed,
negative links decay at rate $\lambda ^{-}$ which we set equal to $\lambda $
for simplicity except when stated otherwise.

This additional mechanism has two effects on the network: firstly, positive
links now disappear at a rate $\lambda+\gamma$ rather than $\lambda$ as
before. Secondly, the rate of introduction of nodes through mutual friends
(the $\xi$ process) is reduced. Since it is the nonlinearity of the $\xi$
process that produces the coexistence region, one might expect this to have
important effects on the size and location of the coexistence region.

\begin{figure}[tbph]
\includegraphics[width=90mm,clip]{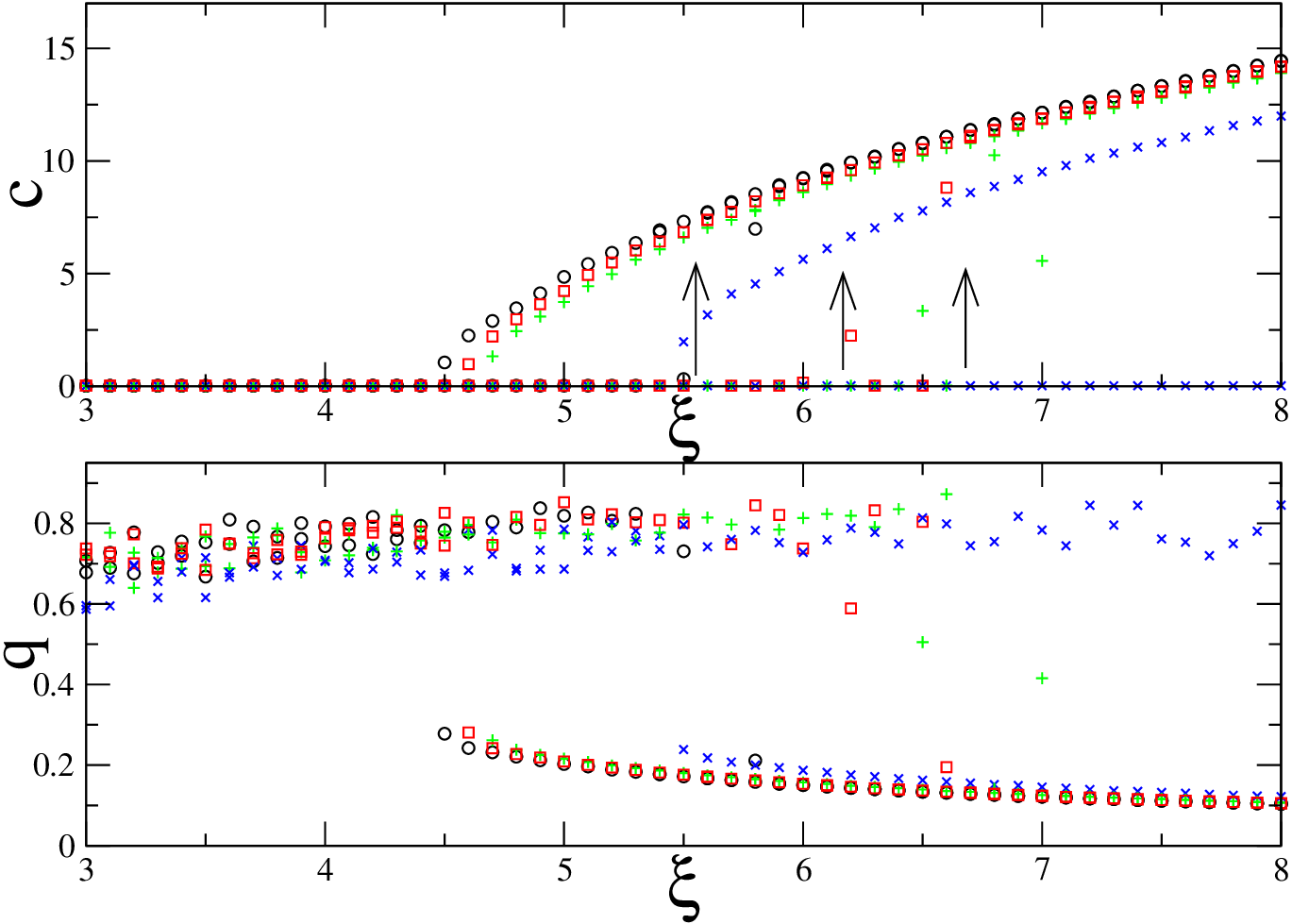}
\caption{Average degree $c$ (top) and clustering coefficient $q$ (bottom)
from numerical simulations with $\protect\eta /\protect\lambda =0.01$ for
populations of size $n=1000$. Original case $\protect\gamma=0$ (black
circle), $\protect\gamma=0.01$ (red square), $\protect\gamma=0.02$ (green
plus), and $\protect\gamma=0.1$ (blue cross). 
The arrows indicate the approximate locations of $\xi_2$ for each value of $\gamma$.
\label{gammafig_vsxi}}
\end{figure}

Figure \ref{gammafig_vsxi} shows plots for four values of $\gamma $. As $%
\gamma $ is increased, the value $\xi _{2}$ above which the low connected
state becomes unstable increases markedly -- indeed for $\gamma =0.1$ the
value of $\xi $ at which the transition occurs (for the times of $3000+3000$
studied here) is around $\xi =20$. Also, the average degree of the network
in the connected region decreases and the value $\xi _{1}$ below which the
connected region collapses moves slightly up. The overall effect is that the
coexistence region gets larger and moves slightly to higher values of $\xi $
when $\gamma $ increases.

More dramatic effect occurs for large values of $\gamma$. Figure \ref%
{figriseandcrash} shows that the system may enter into a regime where the
network undergoes successive rises and crashes due to the spread of
animosity. This behavior also sets in if negative links are much more stable
than positive ones ($\lambda^-\ll \lambda$, lower panel of Fig. \ref%
{figriseandcrash}). Then once a connected society is formed, its network of
relationships gets slowly poisoned with long lasting negative links, which
inhibit the formation of other positive links.

We believe that the occurrence of such time-dependent behaviours is
intimately related to the phase coexistence of the original system. Here the
low connectivity state is unstable, over some mean waiting time, to the
formation of the highly connected state. However the highly connected state
is also not stable once a sufficiently large number of links have turned to
negative links. The system thus alternates between the two states, but not
in a periodic manner due to the stochastic nature of the process.

\begin{figure}[tbph]
%\vspace*{60mm} %
\includegraphics[width=90mm,clip]{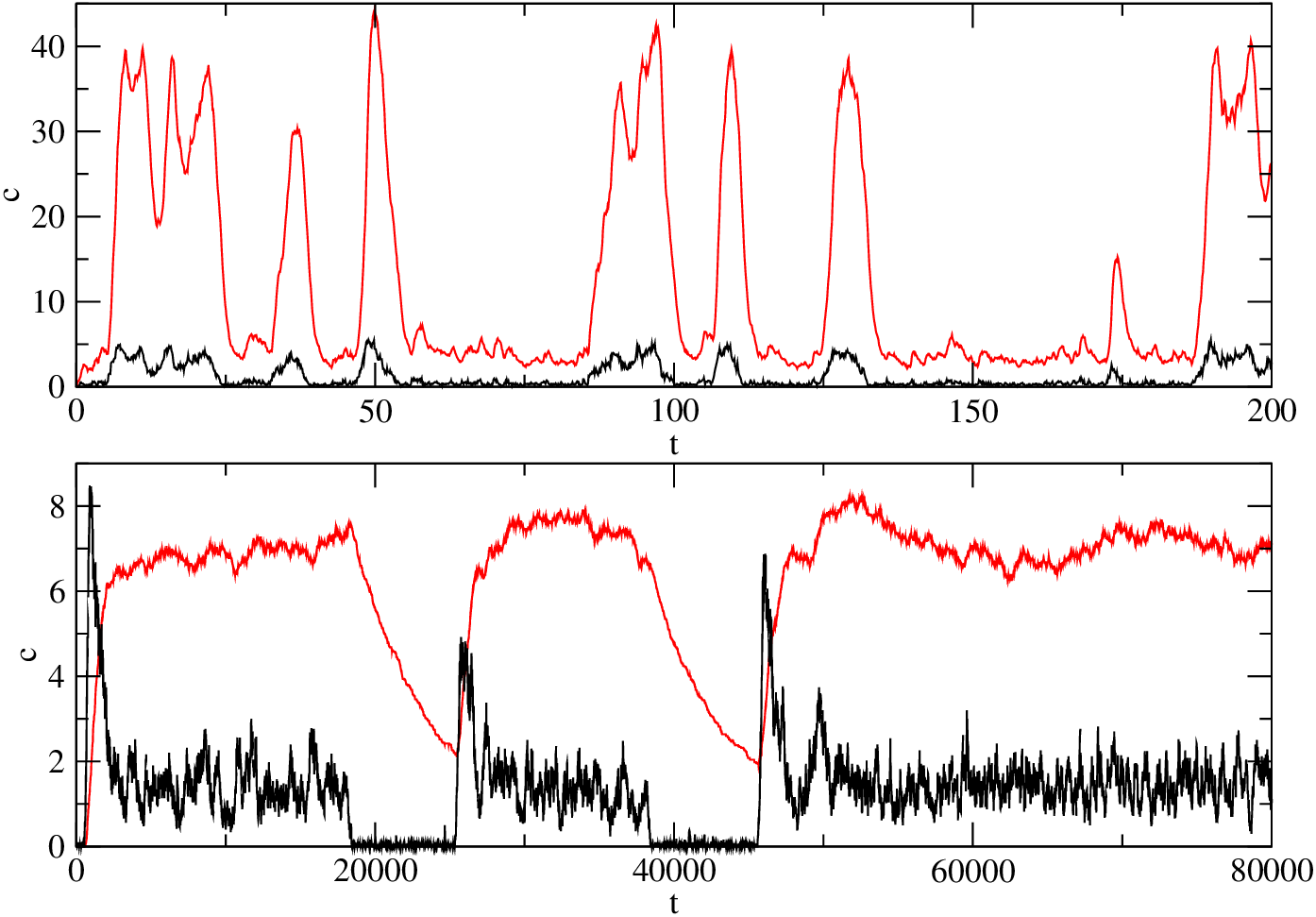}
\caption{Average positive degree $c$ (black) and negative degree (red)
plotted against time. From numerical simulations with populations of size $%
n=100$. For the upper graph the parameters are: $\protect\lambda=\protect%
\lambda^-=1$, $\protect\eta=1$, $\protect\gamma=10$, $\protect\xi=400$.  For
the lower graph the parameters are: $\protect\lambda=1$, $\protect\lambda%
^-=2\times10^{-4}$, $\protect\eta=10^{-2}$, $\protect\gamma=10^{-3}$, $%
\protect\xi=8$. Note that in this case $\protect\lambda \gg \protect\lambda^-
$. }
\label{figriseandcrash}
\end{figure}

\section{The effect of geometry}

We now consider another important effect not considered in the original
model, that of the physical space in which the agents live. We introduce a
modified version of the model which accounts for the fact that agents
embedded in space are more likely to make random acquaintances with other
agents who are geographically near to them.

We modify the original model in the following way: We embed the agents on a
one-dimensional periodic lattice of length $n$, with agent $i$ being placed
at a distance $i$ from the origin. When creating long-range links (the $\eta 
$ process), we select site $i$ at random and then site $j$ with a
probability $P(d_{ij})$ which decays with the distance $d_{ij}$ between $i$
and $j$ on the lattice.\footnote{%
Notice that because of periodic boundary conditions, if $i<j$ then $%
d_{ij}=\min (j-i,i-j+L)$.} We studied distributions of the form $P(d)\propto
d^{-\alpha }$ ($\alpha >0$) decays with distance.

\begin{figure}
\begin{minipage}[t]{8.5cm}%{4.5cm}
\includegraphics[width=0.9\textwidth,clip]{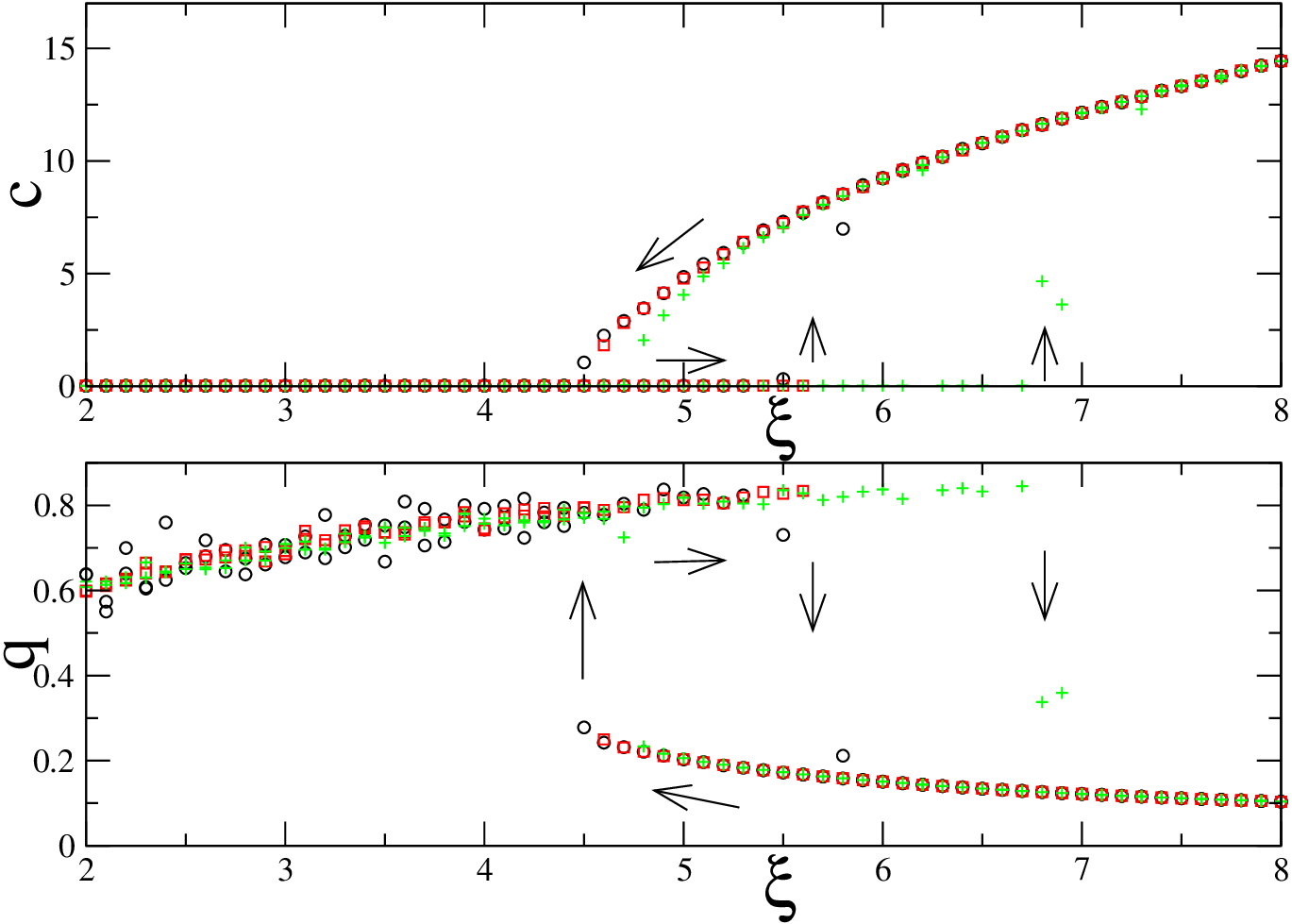} 
\end{minipage}
\hfill
\begin{minipage}[t]{8.5cm}%{4.5cm}
\includegraphics[width=0.9\textwidth,clip]{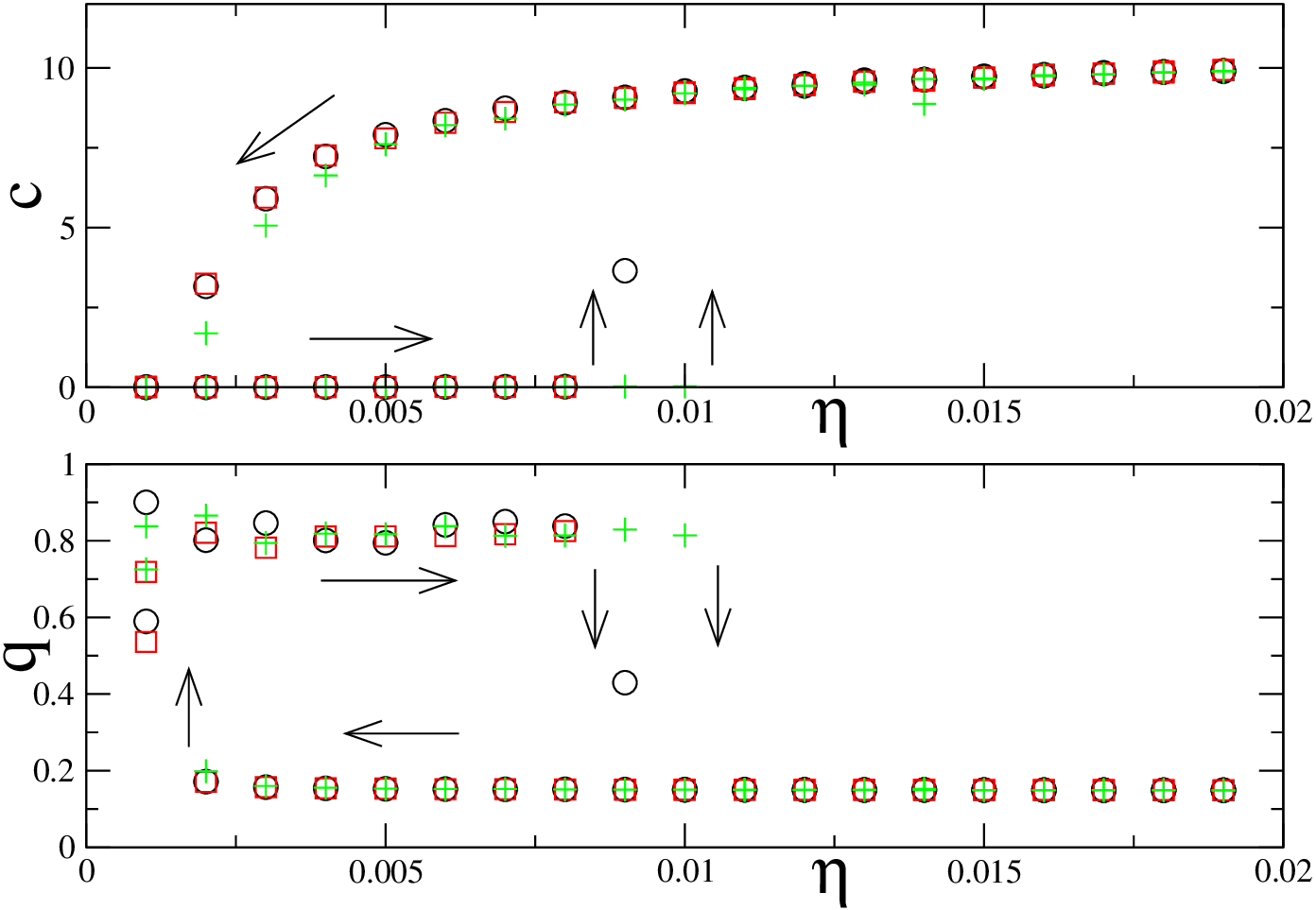} \
\end{minipage}
\hfill
\caption{
Average degree $c$ (top) and clustering coefficient $q$ (bottom) from numerical simulations for populations of size $n=1000$.
The left-hand plots show results for $\protect\eta /\protect\lambda=0.01$, plotted against $\xi$.
The right-hand plots show results for $\protect\xi /\protect\lambda=6$, plotted against $\eta$.
The points are: original case $\alpha=0$ (black circle), $\alpha=1$ (red square), $\alpha=2$ (green plus).
The arrows indicate the points at which transitions occur and the directions in which the system moves within the hysteretic region. 
Notice that the coexistence region is extended for $\alpha=2$.
\label{distvs}
}\end{figure}

Notice that the local search process $\xi $ can only connect members of a
community of already connected agents. It is only by the $\eta $ process
that such a community can reach agents further away. Hence we expect that a
sharp decay of $P(d)$ with distance has strong effects on the $\eta $ process, which is
the limiting factor in the nucleation of a dense network,\footnote{%
Furthermore, this modification has also the effect of reducing the rate at
which links are formed by the $\eta $ process. This is because links between
close agents are more likely to exist already and cannot be added again.}
thus increasing the stability of the low density phase. Figures \ref{distvs} 
confirm this expectation for the case $%
P(d)\sim d^{-\alpha }$ with $\alpha =1$,$2$. The main change occurs for $%
\alpha =2$ where the stability of the low connectivity phase and hence the
coexistence region is significantly extended. Notice also that well inside the
dense network phase there is no significant effect. This confirms that this
phase is sustained by the local search process alone: once a global network
spanning the whole system is formed, geometry has no effect.

%%%%%%%%%%%%%%%%%%%%%%%%%%%%%%%%%%%%%%%%%%%%%%%%
%% BACKMATTER
%%%%%%%%%%%%%%%%%%%%%%%%%%%%%%%%%%%%%%%%%%%%%%%%

\begin{acknowledgments}
  Work supported in part by the European Community's 
Human Potential Programme under contract HPRN-CT-2002-00319, 
STIPCO, and by EVERGROW, integrated project No. 1935 in the
complex systems initiative of the Future and Emerging Technologies
directorate of the IST Priority, EU Sixth Framework.
\end{acknowledgments}

%%%%%%%%%%%%%%%%%%%%%%%%%%%%%%%%%%%%%%%%%%%%%%%%
%% You may have to change the BibTeX style below, depending on your
%% setup or preferences.
%%
%% If the bibliography is produced without BibTeX comment out the
%% following lines and see the aipguide.pdf for further information.
%%
%% For The AIP proceedings layouts use either
%%%%%%%%%%%%%%%%%%%%%%%%%%%%%%%%%%%%%%%%%%%%

% if natbib is available
%\bibliographystyle{aipprocl} % if natbib is missing

%%%%%%%%%%%%%%%%%%%%%%%%%%%%%%%%%%%%%%%%%%%
%% You probably want to use your own bibtex database here
%%%%%%%%%%%%%%%%%%%%%%%%%%%%%%%%%%%%%%%%%%%
\bibliographystyle{aipproc}
\bibliography{emv}

%%%%%%%%%%%%%%%%%%%%%%%%%%%%%%%%%%%%%%%%%%%
%% Just a reminder that you may have to run bibtex
%% All of it up to \end{document} can be removed
%% if you don't like the warning.
%%%%%%%%%%%%%%%%%%%%%%%%%%%%%%%%%%%%%%%%%%%
\IfFileExists{\jobname.bbl}{}  {\typeout{}  %
\typeout{******************************************}  \typeout{** Please run
"bibtex \jobname" to obtain}  \typeout{** the bibliography and then re-run
LaTeX}  \typeout{** twice to fix the references!}  %
\typeout{******************************************}  \typeout{}  }

\end{document}